\documentclass{kapproc}

\usepackage{bm}

\DeclareFontFamily{OT1}{rsfs}{}
\DeclareFontShape{OT1}{rsfs}{m}{n}{
                 <-7> rsfs5 <7-10> rsfs7 <10-> rsfs10}{}
\DeclareMathAlphabet{\mycal}{OT1}{rsfs}{m}{n}

\normallatexbib

\begin{document}

\articletitle{Improving the ``No--Hair'' Theorem for the Proca
Field}

\author{Eloy Ay\'on--Beato}
\affil{Departamento~de~F\'{\i}sica,%
~Centro~de~Investigaci\'{o}n~y~Estudios~Avanzados~del~IPN\\
Apdo.\ Postal 14--740, 07000 M\'exico DF, MEXICO}
\email{ayon@fis.cinvestav.mx}

\chaptitlerunninghead{Improving the ``No--Hair'' Theorem for the
Proca Field}

\begin{abstract}
This paper reconsider the problem of a Proca field in the exterior
of a static black hole. The original Bekenstein's demonstration on
the vanishing of this field, based on an integral identity, is
improved by using more natural arguments at the event horizon. In
particular, the use of the so--called \emph{standard} integration
measure in the horizon is fully justified. Accordingly, the
horizon contribution to the Bekenstein integral identity is more
involved and its vanishing can be only established using the
related Einstein equations. With the new reasoning the
``no--hair'' theorem for the Proca field now rest on better
founded grounds.
\end{abstract}


\section{\label{sec:int} Introduction}

Massive fields are forbidden in the exterior of any stationary
black hole. Such statement rest in the fact that the strongest
version of the ``no--hair'' conjecture establishes that a
stationary black hole is uniquely determined by global charges
(conserved Gauss--like surface integrals at spatial infinity)
\cite{Bizon94}, but massive fields exponentially fall--off at
infinity and made no contributions to the corresponding surface
integrals; there are no global charges associated with them. This
kind of unobserved--from--infinity configuration is what is called
``hair'' in the literature.

The first results explicitly showing the nonexistence of massive
``hair'' were due to Bekenstein who studied massive scalar fields,
Proca--massive spin--1 fields, and massive spin--2 fields
\cite{Bekens72}. This paper intents to improve the original
demonstration of Bekenstein for Proca fields in the presence of
static black holes. Such demonstration depends on an integral
identity only built from matter field equations; no Einstein
equation is used. The fundamental changes introduced in the proof
are related to the arguments concerning the event horizon, which
are usually the more involved. On the one hand, an essentially
appropriate integration measure is introduced on the event
horizon, which is a degenerate hypersurface. As a consequence the
horizon contribution to the cited integral identity is more
elaborate. On the other hand, the Bekenstein proofs are usually
announced as independent of the particular metric theory of
gravity (see \cite{Bekens98}); due to they involve only  matter
field equations. However, we shall show that in order to vanishing
properly the horizon contributions to the Bekenstein's identity,
it is imperative also the use of Einstein's equations. We would
like to point out that the appropriate justification of the
``no--hair'' conjecture for massive vector fields has been a
useful tool for excluding the existence of new black hole
configurations from very complicated theories as metric--affine
gravity, where a relevant sector of this theory reduces to an
effective Einstein--Proca system \cite{AyonGMQ00,ObukhovV99&T00}.
Secondly, but not least important, the methods developed here has
also been a start point in the exclusion of ``hair'' for more
complex system where the mass terms appears dynamically through
spontaneous symmetry breaking \cite{Ayon00}.

In the following Sec.~\ref{sec:Proca} the Einstein--Proca system
is introduced and the consequences of considering this system in
the exterior of a static black hole are highlighted. In
Sec.~\ref{sec:theo} the ``no--hair'' theorem for the Proca field
is established using the Bekenstein argument but with a different
integration measure in the horizon, and helping us of the Einstein
equations in the reasoning. Section \ref{sec:con} is devoted to
the relevant conclusions. The final Appendix is dedicated to
properly justify the use of the standard integration measure in
the horizon mentioned above.

\section{\label{sec:Proca} Proca Fields on Static Spacetimes}

In this section we introduce some fundamental properties of a
Proca field lying in the domain of outer communications
$\ll\!\!\mycal{I}\!\!\gg$ of a static black hole. The
Einstein--Proca action describing such interaction is given by
\begin{equation}
 S = \int\left(\frac1{2\kappa}R - \frac1{16\pi}H_{\mu\nu}H^{\mu\nu}
   - \frac{m^2}{8\pi}B_\mu B^\mu\right) \mathrm{d}v ,
 \label{eq:S}
\end{equation}
where $R$ stands for scalar curvature, and $H_{\mu \nu }\equiv
2\nabla _{[\mu }B_{\nu ]}$ is the field strength of the Proca
field $B_\mu $. From (\ref{eq:S}) the Einstein and Proca equations
are established
\begin{equation}
 \frac{4\pi}\kappa R_{\mu\nu}=H_\mu^{~\alpha}H_{\nu\alpha}
 + m^2B_\mu B_\nu - \frac14g_{\mu\nu}H_{\alpha\beta}H^{\alpha\beta},
 \label{eq:Ein}
\end{equation}
\begin{equation}
 \nabla_\beta H^{\beta\alpha}=m^2B^\alpha .
 \label{eq:Proca}
\end{equation}

In a static black hole the Killing field $\bm{k}$ coincides with
the null generator of the event horizon $\mycal{H}^+$. At the same
time this field  is timelike and hypersurface orthogonal in all
the domain of outer communications $\ll\!\!\mycal{I}\!\!\gg$.
These properties of the Killing field together with the simply
connectedness of $\ll\!\!\mycal{I}\!\!\gg$ \cite{ChrWald95} allow
us to choose a global coordinate system $(t,x^i)$, $i=1,2,3$, in
all $\ll\!\!\mycal{I}\!\!\gg$ \cite{Carter87} such that
$\bm{k}=\bm{\partial/\partial{t}}$ and
\begin{equation}
 \bm{g}=-V\bm{dt}^2+\gamma_{ij}\bm{dx}^i\bm{dx}^j,
 \label{eq:static}
\end{equation}
where $V$ and $\bm{\gamma}$ are $t$--independent, $\bm{\gamma}$ is
positive definite in all $\ll\!\!\mycal{I}\!\!\gg$, and the
function $V$ is positive in all $\ll\!\!\mycal{I}\!\!\gg$ and
vanishes in $\mycal{H}^+$. From (\ref{eq:static}) it can be note
that staticity is equivalent to the existence of a time--reversal
isometry $t\mapsto-t$ in all $\ll\!\!\mycal{I}\!\!\gg$.

We shall assume that the Proca field shares the same symmetries of
the metric; firstly, that it is stationary $\bm{\pounds_kB}=0$.
Secondly, that the staticity of the metric is also extended to the
Proca field $B^\alpha$ and its field equations (\ref{eq:Proca}) in
the sense of requiring they are all invariant under time--reversal
transformations (electromagnetic staticity). The condition of
time--reversal invariance for Proca equations (\ref{eq:Proca})
written in the coordinates of Eq.\ (\ref{eq:static}) demands that
the components $B^t$ and $H^{ti}$ remain unchanged while $B^i$ and
$H^{ij}$ change sign, or the opposite scheme, i.e., $B^t$ and
$H^{ti}$ change sign as long as $B^i$ and $H^{ij}$ remain
unchanged under time reversal \cite{Bekens72}. Therefore, for a
time--reversal invariant Proca field the components $B^i$ and
$H^{ij}$ must vanish in the first case mentioned above, and the
components $B^t$ and $ H^{ti}$ vanish in the second one. Hence,
time--reversal invariance implies the existence of two separated
cases: a purely electric case (I) and a purely magnetic case (II).

\section{\label{sec:theo} The ``No--Proca--Hair'' Theorem}

Now we are ready to proof the ``no--hair'' theorem for the Proca
field and we start by obtaining the corresponding integral
identity mentioned in the introduction. Let
$\mycal{V}\!\subset\ll\!\!\mycal{I}\!\!\gg$ be the open region
bounded by the spacelike hypersurface $\Sigma$, the spacelike
hypersurface $\Sigma^{\prime}$, and the pertinent portions of the
horizon $\mycal{H}^+$, and the spatial infinity $i^o$. The
spacelike hypersurface $\Sigma^{\prime}$ is obtained by shifting
each point of $\Sigma$ a unit parametric value along the integral
curves of the Killing field $\bm{k}$. Multiplying the Proca
equations (\ref{eq:Proca}) by $B_\alpha$ and integrating by parts
over $\mycal{V}$ using the Gauss law, one obtains
\begin{eqnarray}
 \left[ \int_{\Sigma^{\prime}} - \int_\Sigma
 + \int_{\mycal{H}^+\cap\overline{\mycal{V}}}
 + \int_{i^o\cap\overline{\mycal{V}}} \right]
 B_\alpha H^{\beta\alpha}\mathrm{d}\Sigma_\beta &&\nonumber\\
 = \int_{\mycal{V}}\left(\frac12H_{\alpha\beta}H^{\alpha\beta}
 + m^2B_\alpha B^\alpha \right) \mathrm{d}v. &&
 \label{eq:int}
\end{eqnarray}
The boundary integral over $\Sigma^{\prime}$ cancels
out the corresponding one over $\Sigma$, since $\Sigma^{\prime}$
and $\Sigma$ are isometric hypersurfaces taken with reversed
normals in the Gauss law. The boundary integral over the infinity
$i^o\cap\overline{\mycal{V}}$ vanishes by the usual Yukawa
fall--off of massive fields asymptotically.

We will show that the integrand of the remaining boundary integral
at the portion of the horizon
$\mycal{H}^+\cap\overline{\mycal{V}}$ also vanishes. To achieve
this goal we use the standard measure at the horizon
\cite{Zannias},
\begin{equation}
 \mathrm{d}\Sigma_\beta=2n_{[\beta}l_{\mu]}l^\mu\mathrm{d}\sigma,
 \label{eq:meas}
\end{equation}
where $\bm{l}$ is the null generator of the horizon, $\bm{n}$ is
the other future--directed null vector ($n_\mu l^\mu=-1$),
orthogonal to the spacelike cross sections of the horizon, and
$\mathrm{d}\sigma$ is the surface element. We shall justify the
use of the standard measure on the horizon in the final Appendix,
see Eq.~(\ref{eq:null}). By using the quoted measure the horizon
integrand can be written as
\begin{equation}
   B_\alpha H^{\beta\alpha}\mathrm{d}\Sigma_\beta
 = \left(B_\alpha H^{\beta\alpha}l_\beta
 + B_\alpha H^{\beta\alpha}n_\beta\,l_\mu l^\mu\right)
   \mathrm{d}\sigma\,.
 \label{eq:integ}
\end{equation}
In order to show that the last integrand is vanishing it is
sufficient to prove that the quantities inside the parenthesis at
the right--hand side of Eq.\ (\ref{eq:integ}) satisfy the
following conditions: $B_\alpha{H}^{\beta\alpha}l_\beta$ vanishes
and $B_\alpha{H}^{\beta\alpha}n_\beta$ remains bounded at the
horizon. The behavior of these quantities at the horizon can be
established by studying some invariants constructed from the
curvature. Using Einstein equations (\ref{eq:Ein}), we obtain,
${4\pi}R/\kappa=m^2B_\mu B^\mu$ and
\begin{equation}
   \frac{16\pi^2}{\kappa^2}R_{\mu\nu}R^{\mu\nu}
 = 3H^2 + 4I^2 + \left(H-m^2B_\mu B^\mu\right)^2
 + 2m^2H_\mu^{~\alpha}B^\mu H_{\nu\alpha}B^\nu,
 \label{eq:RicciS}
\end{equation}
where $H\equiv H_{\alpha\beta}H^{\alpha\beta}/4$,
$I\equiv{}^{*}\!H_{\alpha\beta}H^{\alpha\beta}/4$, and
${}^{*}\!H_{\alpha\beta}=\eta_{\mu\nu\alpha\beta}H^{\mu\nu}/2$ is
the usual Hodge dual. Since the horizon is a smooth surface
curvature invariants are bounded there, from which it follows
first that $B_\mu B^\mu$ is bounded at the horizon. The last term
in Eq.\ (\ref{eq:RicciS}) is nonnegative in both cases (I) and
(II), the remaining terms are also nonnegative, and consequently
each one is bounded at the horizon, in particular the invariants
$H$ and $I$. Other invariants can be built from the Ricci
curvature (\ref{eq:Ein}) by means of $\bm{l}$ and $\bm{n}$, which
are well--defined smooth vector fields on the horizon. The first
invariant reads
\begin{equation}
   \frac{4\pi}\kappa R_{\mu\nu}n^\mu n^\nu
 = J_\mu J^\mu + m^2(B_\mu n^\mu)^2 - n_\mu n^\mu H\,,
 \label{eq:Riccinn}
\end{equation}
where $J^\mu\equiv H^{\mu\nu}n_\nu$. The last term above vanishes
because the bounded behavior of the invariant $H$. Since $\bm{J}$
is orthogonal to the null vector $\bm{n}$ it must be spacelike or
null ($J_\mu J^\mu\geq0$), therefore each one of the remaining
terms in the right--hand side of Eq.\ (\ref{eq:Riccinn}) must be
bounded. The next invariant to be considered, which vanishes at
the horizon by applying the Raychaudhuri equation to the null
generator \cite{Wald}, reads
\begin{equation}
 0 = \frac{4\pi}\kappa R_{\mu\nu}l^\mu l^\nu
   = D_\mu D^\mu + m^2(B_\mu l^\mu)^2 - l_\mu l^\mu H\,,
 \label{eq:Riccill}
\end{equation}
where $D^\mu \equiv H^{\mu \nu }l_\nu $ is the electric field at
the horizon. Once again the bounded behavior of the invariant $H$
can be used to vanishing the last term of relations
(\ref{eq:Riccill}). The vector $\bm{D}$ is orthogonal to the null
generator $\bm{l}$ hence must be spacelike or null
($D_{\mu}D^\mu\geq0$). Consequently each term on the right--hand
side of Eq.\ (\ref{eq:Riccill}) vanishes independently, which
implies that $B_\mu l^\mu =0$ and that $\bm{D}$ is proportional to
the null generator $\bm{l}$ at the horizon, i.e.,
$\bm{D}=-(D_\alpha{n}^\alpha)\,\bm{l}$. The following relation
arise from the last invariant to be studied
\begin{equation}
 \frac{4\pi}\kappa R_{\mu\nu}l^\mu n^\nu - H
 = (D_\mu n^\mu)^2 + m^2(B_\mu n^\mu)(B_\nu l^\nu),
 \label{eq:Ricciln}
\end{equation}
where it has been used that $\bm{D}=-(D_\alpha n^\alpha)\,\bm{l}$.
Since $B_\mu l^\mu=0$ and $B_\mu n^\mu$ is bounded at the horizon,
it follows that the second term on the right--hand side of Eq.\
(\ref{eq:Ricciln}) vanishes. Therefore, $D_\mu n^\mu$ is bounded
at the horizon as consequence of the bounded behavior of the
related left--hand side in Eq.\ (\ref{eq:Ricciln}).

Summarizing, the study of the horizon behavior of all the above
invariants leads to the following conclusions: the quantities
$D_\mu{n}^\mu$, $B_\mu{n}^\mu$, $B_\mu{B}^\mu$, and $J_\mu J^\mu$
are bounded at the horizon, and the relations $B_\mu\,l^\mu=0$,
and $\bm{D}=-(D_\alpha{n}^\alpha)\,\bm{l}$ are satisfied in the
same region.

Now we are in position to show the fulfillment of the sufficient
conditions for the vanishing of the integrand (\ref{eq:integ})
over the horizon, i.e., that $B_\alpha H^{\beta\alpha}l_\beta$
vanishes and $B_\alpha H^{\beta\alpha}n_\beta$ remains bounded at
the horizon. Using the definition $D^\mu\equiv H^{\mu\nu}l_\nu$
and that $\bm{D}=-(D_\alpha n^\alpha)\,\bm{l}$, we obtain for the
first quantity at the horizon
\begin{equation}
 B_\alpha H^{\beta\alpha}l_\beta = (D_\mu n^\mu)(B_\nu l^\nu) = 0,
 \label{eq:int1v}
\end{equation}
where the vanishing follows from the fact that, as we just
establish, $D_\mu n^\mu$ is bounded and $B_\nu\,l^\nu$ vanishes at
the horizon.

For the second quantity we note that $\bm{B}$ and $\bm{J}$ are
orthogonal to the null vectors $\bm{l}$ and $\bm{n}$,
respectively. Therefore, $\bm{B}$ must be spacelike or
proportional to $\bm{l}$, and $\bm{J}$ must be spacelike or
proportional to $\bm{n}$. Using a null tetrad basis constructed
with $\bm{l}$, $\bm{n}$, and a pair of linearly independent
spacelike vectors, spanning the spacelike cross sections of the
horizon, the $\bm{B}$ and $\bm{J}$ vectors can be written as
\begin{equation}
 \bm{B} = - (B_\alpha n^\alpha)\bm{l} + \bm{B}^{\bot},\qquad
 \bm{J} = - (J_\alpha l^\alpha)\bm{n} + \bm{J}^{\bot},
 \label{eq:BJnt}
\end{equation}
where $\bm{B}^{\bot}$ and $\bm{J}^{\bot}$ are the projections,
orthogonal to $\bm{l}$ and $\bm{n}$, on the spacelike cross
sections of the horizon. Using expressions (\ref{eq:BJnt}) it is
clear that $B_{\mu}B^\mu=B_\mu^{\bot}B^{\bot\mu}$ and
$J_{\mu}J^\mu=J_\mu^{\bot}J^{\bot\mu}$, i.e., the contribution to
these bounded magnitudes comes only from the spacelike sector
orthogonal to $\bm{l}$ and $\bm{n}$. With the help of Eqs.\
(\ref{eq:BJnt}) the other quantity appearing in the integrand
(\ref{eq:integ}) can be written as
\begin{equation}
 B_\alpha H^{\beta\alpha}n_\beta = - B_\alpha J^\alpha
 = - (B_\alpha n^\alpha)(D_\beta n^\beta) - B_\alpha^{\bot}J^{\bot\alpha},
 \label{eq:int2b}
\end{equation}
where the identity $J_\alpha l^\alpha =-D_\alpha n^\alpha $ has
been used. The first term in (\ref{eq:int2b}) is bounded because
$B_\alpha n^\alpha $ and $D_\beta n^\beta $ are bounded. For the
second term we can apply the Schwarz inequality since
$\bm{B}^{\bot}$ and $\bm{J}^{\bot}$ belong to a spacelike
subspace. Thus,
$(B_\alpha^{\bot}J^{\bot\alpha})^2\leq(B_\mu^{\bot}B^{\bot\mu})
(J_\nu^{\bot}J^{\bot\nu})=(B_\mu B^\mu)(J_\nu J^\nu)$ and since
$B_\mu B^\mu$ and $J_\nu J^\nu$ are bounded at the horizon we
conclude that the second term of Eq.\ (\ref{eq:int2b}) is also
bounded.

Finally, the vanishing of the term (\ref{eq:int1v}) and the
bounded behavior of the other term (\ref{eq:int2b}), together with
the null character of $\bm{l}$ at the horizon lead to the
vanishing of the integrand (\ref{eq:integ}) over the event
horizon.

With no contribution from boundary integrals in the identity
(\ref{eq:int}) we shall write the volume integral, using the
coordinates from Eq.\ (\ref{eq:static}), for each one of the
different cases discussed at the beginning of this section.

For the purely electric case (I) we have
\begin{equation}
 \int_{\mycal{V}} - V\left(\frac12\gamma_{ij}H^{ti}H^{tj}
 + m^2(B^t)^2\right) \mathrm{d}v = 0.
 \label{eq:zeroI}
\end{equation}
The nonpositiveness of the above integrand, which is minus the sum
of squared terms, implies that the integral is vanishing only if
$H^{ti}$ and $B^t$ vanish everywhere in $\mycal{V}$, and hence in
all $\ll\!\!\mycal{I}\!\!\gg$.

For the purely magnetic case (II) the volume integral reads as
\begin{equation}
 \int_{\mycal{V}}\left(\frac12\gamma_{ik}\gamma_{jl}H^{kl}H^{ij}
 + m^2B_iB^i\right) \mathrm{d}v = 0,
 \label{eq:zeroII}
\end{equation}
in this case the non--negativeness of the above integrand is
responsible for the vanishing of $H^{ij}$ and $B^i$ in all
$\ll\!\!\mycal{I}\!\!\gg$.

\section{\label{sec:con} Conclusions}

Concluding, we have proved that the Proca field $\bm{B}$ is
trivial in the presence of a static black hole. It must be pointed
that we improve the original proof of Bekenstein on the subject by
using an appropriate integration measure on the event horizon of
the black hole, and also making explicit use of the gravitational
field equations. The vanishing of $\bm{B}$ implies that the action
(\ref{eq:S}) reduces to the Einstein--Hilbert one, for which the
only static black hole is the Schwarzschild solution (see
\cite{Heusler,Chrusciel02} for references on improvements to the
original proofs). The existence of static soliton (particle--like)
configurations can be also excluded using similar arguments, since
the only change in the proof is that in this case the boundary of
the volume $\mycal{V}$ only consists of the isometric surfaces
$\Sigma$ and $\Sigma^{\prime}$, and a portion of the spatial
infinity $i^o$, i.e., there is no interior boundary corresponding
to the event horizon.

\begin{acknowledgments}
The author thanks Alberto Garc\'{\i}a, Alfredo Mac\'{\i}as,
Hernando Quevedo, and Thomas Zannias for useful discussions and
hints. This research was partially supported by the CONACyT Grant
38495E. The author also thanks Isabel Negrete for typing the
manuscript.
\end{acknowledgments}

\chapappendix{\label{sec:app} On the Suitable Integration Measure
of the Horizon}

In this appendix we justify the use of the standard integration
measure (\ref{eq:meas}) in the boundary integrals on the event
horizon. In the derivation of the basic identity (\ref{eq:int}) we
make use of Gauss's law, which is a well--known particular form of
Stokes's theorem
\begin{equation}
   \int_\mycal{V} \bm{d\alpha}
 = \int_{\partial\mycal{V}} \bm{\alpha}
   \qquad\Longrightarrow\qquad
   \int_\mycal{V} \nabla_\beta v^\beta \mathrm{d}v
 = \int_{\partial\mycal{V}} v^\beta \mathrm{d}\Sigma_\beta,
 \label{eq:StokesGauss}
\end{equation}
for some volume $\mycal{V}$ with boundary $\partial\mycal{V}$. The
relation between both theorems rest on that the three--form
$\bm{\alpha}$ is the Hodge dual of the vector field $\bm{v}$
\cite{Wald}, we shall explore such relation in order to find the
horizon integration measure. In the Stokes version we can write
the boundary integrand as $\bm{\alpha}=h\,\bm{\eta_3}$ using that
the three--form $\bm{\alpha}$ must be proportional to the volume
three--form $\bm{\eta_3}$ of the boundary $\partial\mycal{V}$.

For example, in the case of a boundary consisting of non--null
surfaces the induced metric there is nondegenerate, and we can
choose as volume three--form on $\partial\mycal{V}$ the one
associated with the induced metric. It is given by
${\eta_3}_{\alpha\beta\gamma}=\pm{}^{*}\tilde{n}_{\alpha\beta\gamma}
\equiv \pm\eta_{\mu\alpha\beta\gamma}\tilde{n}^\mu$, where
$\bm{\eta}$ is the four--dimensional volume form, $\bm{\tilde{n}}$
is the unit normal to $\partial\mycal{V}$, and we use the plus
sign if $\bm{\tilde{n}}$ is spacelike and the minus one if is
timelike; in both cases the normal is chosen to be ``outward
pointing'' in the volume $\mycal{V}$ in order to keep the
orientation needed in Stokes's theorem \cite{Wald}. For this
election of the boundary volume three--form we have the relation
$\bm{\alpha}={}^{**}\bm{\alpha}=\pm{h}\,{}^{*}\bm{\tilde{n}}$,
from which it follows that $h={}^{*}\alpha^\beta\tilde{n}_\beta$,
and we recover the Gauss form of the boundary integral
\begin{equation}\label{eq:nonull}
   \int_{\partial\mycal{V}} \bm{\alpha}
 = \int_{\partial\mycal{V}} {}^{*}\alpha^\beta
   \tilde{n}_\beta\,\bm{\eta_3}
 = \int_{\partial\mycal{V}} v^\beta \mathrm{d}\Sigma_\beta\,,
\end{equation}
here $v^\beta={}^{*}\alpha^\beta
=\eta^{\mu\nu\gamma\beta}\alpha_{\mu\nu\gamma}/3!$, and the
integration measure at the boundary is the traditional one for
non--null surfaces
$\mathrm{d}\Sigma_\beta=\tilde{n}_\beta\,\bm{\eta_3}
=\tilde{n}_{\beta}\mathrm{d}\sigma$, where $\mathrm{d}\sigma$
stands for the volume element ($\mathrm{d}\sigma=\bm{\eta_3}$)
following the usual notation.

The above situation does not apply to null surfaces, which is our
case of interest when we try to integrate on the horizon. In this
case the induced metric is degenerate, and a priori there is no
natural choice for the volume form. However, in the case of the
event horizon we can use other geometrical objects naturally
defined on it to specify a volume three--form. Let $\bm{l}$ be the
null generator of the horizon, and $\bm{n}$ be the other linearly
independent and future--directed null vector orthogonal to the
spacelike cross sections of the horizon, and normalized in such a
way that $n_{\mu}l^\mu=-1$. For smooth event horizon they are
well--defined smooth vector fields along it. In this case the
volume three--form must not be orthogonal to $\bm{l}$ since such
vector is tangent to the horizon, in fact, their interior product
must coincide with the two--form expanding the area of spacelike
cross sections of horizon which is obviously given by
${}^{*}(\bm{l}\wedge\bm{n})$. Hence, the volume three--form
$\bm{\eta_3}$ must satisfy the relation
${\eta_3}_{\mu\alpha\beta}l^\mu={}^{*}({l}\wedge{n})_{\alpha\beta}$,
using now the identity ${\eta_3}_{\mu\alpha\beta}l^\mu
=-{}^{*}(l\wedge\!{}^{*}\eta_3)_{\alpha\beta}$ \cite{Heusler} we
conclude that a natural election for the volume three--form at the
horizon is $\bm{\eta_3}=-{}^{*}\bm{n}$. Now we can find the
function $h$ inside the boundary integrand; multiplying the
relation
$\alpha_{\alpha\beta\gamma}=-h\,\eta_{\rho\alpha\beta\gamma}n^\rho$
by the three--form
$\left[{}^{*}({l}\wedge{n})\wedge{l}\right]^{\alpha\beta\gamma}
=3\,\eta^{\mu\nu[\alpha\beta}l_{\mu}n_{\nu}l^{\gamma]}$ we obtain
\[
   3\,\alpha_{\alpha\beta\gamma}
   \eta^{\mu\nu\alpha\beta}l_{\mu}n_{\nu}l^{\gamma}
 = -3\,h\,\eta_{\rho\alpha\beta\gamma}n^\rho
   \eta^{\mu\nu\alpha\beta}l_{\mu}n_{\nu}l^{\gamma}
 = 3!\,h,
\]
and expanding the left--hand side above using that
$\bm{\alpha}={}^{**}\bm{\alpha}$ we have finally
\begin{equation}\label{eq:h}
 h = \frac12\eta_{\rho\alpha\beta\gamma}{}^{*}\alpha^\rho
     \eta^{\mu\nu\alpha\beta}l_{\mu}n_{\nu}l^{\gamma}
   = 2 {}^{*}\alpha^{\rho} n_{[\rho}l_{\gamma]}l^{\gamma}.
\end{equation}
Hence the boundary integral can be expressed in the Gauss form as
\begin{equation}\label{eq:null}
   \int_{\partial\mycal{V}} \bm{\alpha}
 = \int_{\partial\mycal{V}}
   {}^{*}\alpha^{\beta}2n_{[\beta}l_{\mu]}l^{\mu}\,\bm{\eta_3}
 = \int_{\partial\mycal{V}} v^\beta \mathrm{d}\Sigma_\beta\,,
\end{equation}
where again $v^\beta={}^{*}\alpha^\beta$, but this time the
boundary integration measure is expressed as
$\mathrm{d}\Sigma_\beta=2n_{[\beta}l_{\mu]}l^{\mu}\mathrm{d}\sigma$,
and we use the standard notation for the volume element
$\mathrm{d}\sigma=\bm{\eta_3}$. This is the boundary integration
measure introduced in Eq.\ (\ref{eq:meas}) for the boundary
integral at the event horizon and also used in previous references
\cite{Zannias,AyonGMQ00,Ayon00}.


\begin{chapthebibliography}{1}

\bibitem{Bizon94} P. Bizo\'n,
\emph{Acta Phys. Polon.} \textbf{B25} (1994) 877.

\bibitem{Bekens72} J.D. Bekenstein,
\emph{Phys. Rev. Lett.} \textbf{28} (1972) 452;
\emph{Phys. Rev.} \textbf{D5} (1972) 1239;
                  \textbf{D5} (1972) 2403.

\bibitem{Bekens98} J.D. Bekenstein,
``Black Holes: Classical Properties, Thermodynamics and Heuristic
Quantization,'' in: \emph{Proceedings of the 9th Brazilian School
of Cosmology and Gravitation}, Rio de Janeiro, Brazil (1998)
\texttt{gr-qc/9808028}.

\bibitem{AyonGMQ00} E. Ay\'on--Beato, A. Garc\'\i a,
                    A. Mac\'\i as and H. Quevedo,
\emph{Phys. Rev.} \textbf{D61} (2000) 084017;
                  \textbf{D64} (2001) 024026.

\bibitem{ObukhovV99&T00} Y.N. Obukhov, E.J. Vlachynsky,
\emph{Annals Phys.} \textbf{8} (1999) 497;
                         M. Toussaint,
\emph{Gen. Rel. Grav.} \textbf{32} (2000) 1689.

\bibitem{Ayon00} E. Ay\'on--Beato,
\emph{Phys. Rev.} \textbf{D62} (2000) 104004.

\bibitem{ChrWald95} P.T. Chru\'sciel and R.M. Wald,
\emph{Class. Quant. Grav.} \textbf{11} (1994) L147.

\bibitem{Carter87} B. Carter,
in: \emph{Gravitation in Astrophysics (Carg\`ese Summer School
1986)}, eds. B. Carter, J.B. Hartle (Plenum, New York 1987).

\bibitem{Zannias} T. Zannias,
\emph{J. Math. Phys.} \textbf{36} (1995) 6970;
                      \textbf{39} (1998) 6651.

\bibitem{Wald} R.M. Wald, \emph{General Relativity}
(Univ. of Chicago Press, Chicago 1984).

\bibitem{Heusler} M. Heusler,
\emph{Black Hole Uniqueness Theorems}
(Cambridge Univ. Press, Cambridge 1996);
\emph{Living Rev. Rel.} \textbf{1}, 1998--6,\\
\texttt{http://www.livingreviews.org/Articles/Volume1/1998-6heusler}.

\bibitem{Chrusciel02} P.T. Chru\'{s}ciel,
``Black Holes,'' \texttt{gr-qc/0201053}.

\end{chapthebibliography}
\end{document}